\documentclass[11pt, a4paper]{article}
\pdfoutput=1
\usepackage{jcappub}

\usepackage{booktabs}
\usepackage{multirow}
\usepackage{amssymb}
\usepackage[export]{adjustbox}
\usepackage{floatrow}
\newfloatcommand{capbtabbox}{table}[][3.5in]
\newfloatcommand{capbfigbox}{figure}[][2.3in]
\usepackage{blindtext}
\usepackage{amsmath}
\usepackage{booktabs}
\usepackage{aas_macros}

\def\lsim{\mathrel{\raise.3ex\hbox{$<$\kern-.75em\lower1ex\hbox{$\sim$}}}}
\def\gsim{\mathrel{\raise.3ex\hbox{$>$\kern-.75em\lower1ex\hbox{$\sim$}}}}

\begin{document}

\title{$\gamma$-Cascade: A Simple Program to Compute Cosmological Gamma-Ray Propagation}

\author{Carlos Blanco}\note{ORCID: http://orcid.org/0000-0001-8971-834X}
\emailAdd{carlosblanco2718@uchicago.edu}

\affiliation{University of Chicago, Department of Physics, Chicago, IL USA}
\affiliation{University of Chicago, Kavli Institute for Cosmological Physics, Chicago, IL USA}

\abstract{Modeling electromagnetic cascades during gamma-ray transport is important in many applications within astrophysics. This document introduces $\gamma$-Cascade, a publicly available Mathematica package which allows users to calculate observed gamma-ray fluxes from point sources as well as from a distribution of sources. $\gamma$-Cascade semi-analytically computes the effects of electromagnetic interactions during gamma-ray transport.}

\maketitle

\section{Introduction and Structure of $\gamma$-Cascade}
While the universe is largely transparent to gamma rays with energies below about a hundred GeV, interactions occurring at TeV energies and above produce a substantial cascade contribution to the observed spectrum from cosmological sources~\cite{Murase2012a,Berezinsky:2016feh,Venters:2010bq}. Observations from Fermi-LAT, HESS, MAGIC, and VERITAS span energies between hundreds of MeV to several TeV. In the near future, HAWC and CTA will further extend the relevant energy range up to several hundred TeV. Accurately modeling the evolution of spectra during transport in and above this energy range is necessary to understand the physics at the source of the radiation.  Implementing the physics model laid out in this paper, gamma-ray observables have been calculated in studies involving hadronic interactions in radio galaxies~\cite{Blanco:2017bgl} as well as PeV-scale dark matter in dwarf galaxies and galaxy clusters~\cite{blanco2017novel}.

This document introduces $\gamma$-Cascade, a Mathematica package which models gamma-ray propagation over cosmological distances using a semi-analytic strategy, taking into account electromagnetic cascade development, synchrotron cooling, and cosmological expansion. $\gamma$-Cascade allows users to propagate injected spectra from point sources as well as from a distribution of sources and export observed spectra. The energy range which is taken in to account by this package is $10^{-1}$ GeV to $10^{12}$ GeV, which covers the range in which electromagnetic cascades evolve, but remains under the energy at which double and triple pair production becomes important. Higher order processes like double and triple pair production come to dominate the energy attenuation above $E_{\gamma}\approx10^{13}\text{GeV}$\cite{sigl2017astroparticle}.  Furthermore, since Mathematica packages are inherently open-source, the user has direct control over functions, variables, and details of the physics model laid out herein. 

Popular publicly available software such as GALPROP~\cite{strong2009galprop,vladimirov2010galprop} and CRPropa~\cite{batista2016crpropa} provide a variety of propagation tools focusing on cosmic rays. A complementary solution is also offered by ELMAG ~\cite{kachelriess2011elmag}, whose Monte Carlo approach produces results in agreement with those of $\gamma$-Cascade. Analytical approaches to the problem can be found in refs.~\cite{lee58propagation,berezinsky2016high,kalashev2015simulations}.

The physics model used by $\gamma$-Cascade is encapsulated in tabulated libraries. The $\gamma$-Cascade $\texttt{.wl}$ package file is distributed with pre-calculated libraries as well as a tutorial Mathematica notebook which serves as a walk-through of $\gamma$-Cascade assuming little to no knowledge of Mathematica. It is possible for the user to implement modifications to the physics model by exporting new libraries.

The $\gamma$-Cascade package and supporting material can be found at \url{https://github.com/GammaCascade/GCascade}

\section{Gamma-Ray Attenuation and Cascades}

High energy gamma rays interact with radiation fields during transport to produce electron-positron pairs. These charged leptons then up-scatter low energy photons, contributing to the spectrum of propagating gamma rays. At the same time, the charged leptons are deflected by magnetic fields and lose energy through synchrotron emission. This cascade cycle continuously shapes the spectrum of propagating high energy photons.

The optical depth for pair-production in an isotropic radiation field is given by:
\begin{eqnarray}
\tau_{\gamma \gamma}(E_{\gamma}) & = & \intop\intop\sigma_{\gamma\gamma}(E_{\gamma},\epsilon)\frac{dn(\epsilon,r)}{d\epsilon}d\epsilon dr\\
 & \approx & l\int\sigma_{\gamma\gamma}(E_{\gamma},\epsilon)\frac{dn(\epsilon)}{d\epsilon}d\epsilon,\nonumber 
\end{eqnarray}
where $E_{\gamma}$ is the energy of the incoming gamma ray, $\epsilon$ is
the energy of a target photon, and $dn(\epsilon,r)/d\epsilon$ is the
differential number density of the target photons at location, $r$. The second line is applicable to the case of homogeneous radiation fields where $l$ is the distance traversed. $\sigma_{\gamma\gamma}$ is the total pair-production cross section which is approximated to within 3$\%$ accuracy by the following expression~\cite{aharonian1983}
\begin{eqnarray}
\sigma_{\gamma\gamma}(E_{\gamma},\epsilon) & = & \frac{3\sigma_{T}}{2s^{2}} \bigg[ \left(s-1+\frac{1}{2s}-\frac{\ln s}{2}+\ln2\right) \,  \ln\left(\sqrt{s}+\sqrt{s-1}\right) \\
 & + & \frac{\left(\ln s\right)^{2}}{8}-\frac{\left(\ln\left(\sqrt{s}+\sqrt{s-1}\right)\right)^{2}}{2}+\frac{\ln2 \, \ln s}{2}-\sqrt{s^{2}-s} \bigg],\nonumber 
\end{eqnarray}
where $s=E_{\gamma}\epsilon/m_{e}^{2}$ and $\sigma_{T}$ is the Thomson
cross section.

The differential spectrum of electrons and positrons generated in these interactions is given as follows:
\begin{equation}
\frac{dN_e}{dE_{e}}(E_e)=l\iint\frac{dN_{\gamma}}{dE_{\gamma}}\left(E_{\gamma}\right)\frac{dn}{d\epsilon}(\epsilon)\frac{d\sigma_{\gamma\gamma}}{dE_{e}}(\epsilon,E_{\gamma},E_{e}) \, d\epsilon \,dE_{\gamma}, 
\end{equation}
where $dN_{\gamma}/dE_{\gamma}$ is the spectrum of gamma rays injected from the source and $d\sigma_{\gamma\gamma}/dE_{e}$ is the differential cross section for pair production, given by~\cite{aharonian1983}:
\begin{eqnarray}
\frac{d\sigma_{\gamma\gamma}(\epsilon,E_{\gamma},E_{e})}{dE_{e}} & = & \frac{3\sigma_{T}m_{e}^{4}}{32\epsilon^{2}E_{\gamma}^{3}}\left[\frac{4E_{\gamma}^{2}}{\left(E_{\gamma}-E_{e}\right)E_{e}}\ln\left(\frac{4\epsilon E_{e}\left(E_{\gamma}-E_{e}\right)}{m_{e}^{2}E_{\gamma}}\right)-\frac{8\epsilon E_{\gamma}}{m_{e}^{2}}\right.\\
 & + & \left.\left(\frac{2E_{\gamma}^{2}\left(2\epsilon E_{\gamma}-m_{e}^{2}\right)}{\left(E_{\gamma}-E_{e}\right)E_{e}m_{e}^{2}}\right)-\left(1-\frac{m_{e}^{2}}{\epsilon E_{\gamma}}\right)\frac{E_{\gamma}^{4}}{\left(E_{\gamma}-E_{e}\right)^{2}E_{e}^{2}}\right].\nonumber 
\end{eqnarray}

Once created, lepton pairs very quickly inverse Compton scatter (ICS) with low energy photons from background radiation fields producing gamma rays. At the same time, these charged leptons experience synchrotron cooling. During a single interaction, a lepton of energy, $E_{e}$, loses a fraction of energy, $\Delta E_e$, and produces the following gamma-ray ICS spectrum:
\begin{eqnarray}
\frac{dN_{\gamma}}{dE_{\gamma}}\left(E_{\gamma},E_{e}\right)_{\Delta E_{e}} & = & A\left(E_{e},\Delta E_{e}\right)f_{ICS}\left(E_{e}\right)l_{e}\int\frac{dn}{d\epsilon}(\epsilon)\frac{d\sigma_{ICS}}{dE_{\gamma}}(\epsilon,E_{\gamma},E_{e})d\epsilon,
\end{eqnarray}
where $A$ is a normalization factor which is set by $\Delta E_{e}=\int E_{\gamma}\left( dN_{\gamma}/dE_{\gamma} \right) dE_{\gamma}$, and $f_{ICS}$ is the fraction of energy which goes into ICS (the remaining fraction, $1-f_{ICS}$, goes into synchrotron emission). Note that inverse Compton scattering is Klein-Nishina suppressed at the highest energies, which allows the leptons to lose their energy primarily through synchrotron emission.

The differential cross section for ICS is given by the following~\citep{aharonian1981}: 
\begin{eqnarray}
\frac{d\sigma_{ICS}}{dE_{\gamma}}(\epsilon,E_{\gamma},E_{e}) & = & \frac{3\sigma_{T}m_{e}^{2}}{4\epsilon E_{e}^{2}}\left[1+\left(\frac{z^{2}}{2\left(1-z\right)}\right)+\left(\frac{z}{\beta\left(1-z\right)}\right)-\left(\frac{2z^{2}}{\beta^{2}\left(1-z\right)}\right)\right.\\
 & - & \left(\frac{z^{3}}{2\beta\left(1-z\right)^{2}}\right)-\left(\frac{2z}{\beta\left(1-z\right)}\right)\ln\left(\frac{\beta\left(1-z\right)}{z}\right) \bigg], \nonumber 
\end{eqnarray}
where $\beta\equiv4\epsilon E_{e}/m_{e}^{2}$ and $z\equiv E_{\gamma}/E_{e}$.
The total spectrum, $dN_{\gamma}/dE_{\gamma}$, of photons in a cascade created by an electron of energy, $E_{e}$, is then calculated by taking a sum over the spectra generated as the charged lepton loses its energy through successive scatterings:
\begin{equation}
\frac{dN_{\gamma}}{dE_{\gamma}}\left(E_{\gamma},E_{e}\right)=\sum_{\Delta E_{e}}\frac{dN_{\gamma}}{dE_{\gamma}}(E_{\gamma},E_{e})_{\Delta E_{e}}
\end{equation}
It is then possible to obtain the total cascade spectrum generated by an incoming gamma-ray spectrum by combining equations 2.7, 2.5, and 2.3:
\begin{eqnarray}
\frac{dN_{\gamma}}{dE_{\gamma}}\left(E_{\gamma},E_{\gamma'}\right) & = & \int dE_{e}\frac{dN_{e}}{dE_{e}}\left(E_{\gamma'},E_{e}\right)\left[\sum_{\Delta E_{e}}\frac{dN_{\gamma}}{dE_{\gamma}}(E_{\gamma},E_{e})_{\Delta E_{e}}\right]\\
 & = & \int \frac{dN_{\gamma'}}{dE_{\gamma'}}\left(E_{\gamma'}\right)\cdot M\left(E_{\gamma},E_{\gamma'}\right)dE_{\gamma'},\nonumber
\end{eqnarray}
where $E_{\gamma'}$ and $E_{\gamma}$ are the primary and secondary (cascaded) gamma-ray energies, respectively. In the second line, the integrals over both background photon energies, $\epsilon$ and $\epsilon'$, and the sum over integrals of electron energy, $E_{e}$, have been condensed into $M\left(E_{\gamma},E_{\gamma'}\right)$ which is only a function of the interacting gamma-ray energy and the cascaded gamma-ray energy. $M$ can be tabulated from a given background radiation field model and intergalactic magnetic field. $\gamma$-Cascade uses a radiation field model consisting of the CMB and the extragalactic background light (EBL) model defined by Dominguez~\cite{Dominguez:2010bv} with a nominal magnetic field of $10^{-13}\,$G. However, the magnetic field can be set by the user. This choice of EBL model is motivated by the agreement of proagation models using this EBL model with gamma-ray observations in the energy range of hundreds of MeV to tens of GeV~\cite{dominguez2015spectral,moralejo2017measurement,ajello20173fhl}.

 In a propagation step, $l$,  small enough for the optical depth to be much less than one, we define a function, $F$, which acts on the injected gamma-ray spectrum and produces the resulting spectrum after a single cascade cycle. During a cascade cycle, the pair production spectrum and resulting secondary gamma-ray spectrum are calculated once. $F$ is defined as follows:
\begin{equation}
F \left( \frac{dN_{\gamma'}}{dE_{\gamma'}}\right)=\int \frac{dN_{\gamma'}}{dE_{\gamma'}}\left(E_{\gamma'}\right)_{inj}\cdot M\left(E_{\gamma},E_{\gamma'}\right)dE_{\gamma'}.
\end{equation}
Propagating the injected spectrum, $\left(\frac{dN_{\gamma'}}{dE_{\gamma'}}\right)_{inj}$, through a distance, $D$, is equivalent to repeatedly acting on the injected spectrum with $F$. The propagation function, $P$, is then as follows:
\begin{equation}
P \left(D, \left(\frac{dN_{\gamma'}}{dE_{\gamma'}}\right)_{inj}\right)= F^{n} \left( \left(\frac{dN_{\gamma'}}{dE_{\gamma'}}\right)_{inj}\right),
\end{equation}
where $n$ is the number of times that $F$ must be composed with itself, i.e. $nl=D$.

\section{Cosmological Propagation}
Cosmological propagation in $\gamma$-Cascade is simulated in one of two scenarios:
\begin{itemize}
\item Gamma rays injected by a point source.
\item Gamma rays injected by a distribution of sources creating a diffuse background.
\end{itemize}

Point source propagation is the simplest case. Suppose a source injects a differential flux spectrum, $\left( dN_{\gamma}/dE_{\gamma} \right)_{inj}$, with units of $ \text{GeV}^{-1}\,\text{s}^{-1}$, such that the total luminosity of the source is $L=\int E_{\gamma} \left( dN_{\gamma}/dE_{\gamma} \right)_{inj} dE_{\gamma}$. The observed flux is then given by the following: 
\begin{equation}
\frac{dN_{\gamma}}{dE_{\gamma}}=P\left(D_{c}(z),\frac{dN_{\gamma}}{dE_{\gamma}}\left(E_{\gamma}\left(1+z\right)\right))_{inj}\right)\cdot \frac{\left(1+z\right)^2}{4\pi D_{L}^{2}},
\end{equation}
where $D_{L}=(1+z)D_{c}(z)$ is the luminosity distance and $D_{c}(z)$ is the comoving distance assuming a universe with zero curvature. The observed spectrum has units of $ \text{GeV}^{-1}\,\text{s}^{-1}\,\text{cm}^{-2}$.

Diffuse background calculation is a simple generalization of the point source case. Suppose a class of sources has an intrinsic differential flux spectrum, $\left( dN_{\gamma}/dE_{\gamma} \right)_{inj}$ with units of $\text{GeV}^{-1}\,\text{s}^{-1}$. For a cosmological history given by $\left( dN_{\gamma}/dE_{\gamma}dV_{c} \right)_{inj}$, in units of differential flux per unit comoving volume, the observed diffuse flux, in units of $ \text{GeV}^{-1}\,\text{s}^{-1}\,\text{cm}^{-2}\,\text{sr}^{-1}$, is the given by the following:
\begin{equation}
\frac{dN_{\gamma}}{dE_{\gamma}d\Omega}=\int P\left(D_{c}(z),\frac{dN_{\gamma}}{dE_{\gamma}}\left(E_{\gamma}\left(1+z\right)\right)_{inj}\right)\cdot \frac{D_{H} \rho (z)dz }{4\pi E(z)},
\end{equation}
where $D_{H} = c/H_{0}$ is the Hubble distance and $E(z)=\sqrt{\Omega_{M}(1+z)^{3}+\Omega_{\Lambda}}$ for values of $\Omega_{M} = 0.308$, $\Omega_{\Lambda} = 0.691$, and $H_{0}=67.7\,\left( \text{km}\,\text{s}^{-1}\,\text{Mpc}^{-1} \right)$. Note that the cosmological history of the class of sources is assumed to be separable, i.e. the intrinsic differential flux does not change with $z$. More explicitly, $\gamma$-Cascade handles one class of sources at a time with a cosmological history given by the following:
\begin{equation}
\left( \frac{dN_{\gamma}}{dE_{\gamma}dV_{c}} \right)_{inj} = \left( \frac{dN_{\gamma}}{dE_{\gamma}} \right)_{inj} \cdot \rho (z),
\end{equation}
where $\rho(z)$ is the density per unit comoving volume, referred to as the luminosity function.

\section{$\gamma$-Cascade Functions}

$\gamma$-Cascade provides four gamma-ray propagation functions: $\texttt{GCascadeAttenuate}$,\\ $\texttt{GCascadePoint}$, $\texttt{GCascadeDiffuseConstant}$ and $\texttt{GCascadeDiffuse}$. These functions calculate gamma-ray propagation in the energy range of $10^{-1}$ - $10^{12}$ GeV. Additionally, $\gamma$-Cascade makes available a variety of useful arrays which tabulate energies and distances in order to make formatting inputs a straight-forward task. The functions which handle gamma-ray propagation are as follows:
\begin{itemize}

\item $\texttt{GCascadeAttenuate[injected spectrum, source distance (z)]}$ is a function which produces an attenuated differential spectrum  without including cascade evolution but taking into account cosmological energy shifts. The injected spectrum should be formatted as an array of values of differential flux, $dN_{\gamma}/dE_{\gamma }$ $\left(\text{GeV}^{-1}\, \text{s}^{-1}\right)$, evaluated at the energies given by the array, $\texttt{energies}$. The result obtained using this function is a list of values of differential flux, $dN_{\gamma}/dE_{\gamma }$ $\left(\text{GeV}^{-1}\,\text{cm}^{-2}\, \text{s}^{-1}\right)$, evaluated at the energies given by the array, $\texttt{energies}$.


\item $\texttt{GCascadePoint[injected spectrum, source distance (z)]}$ is a function which produces the observed differential spectrum taking into account account cosmological energy shifts and electromagnetic cascade evolution. The injected differential spectrum should be formatted as an array of values of differential flux, $dN_{\gamma}/dE_{\gamma }$ $\left( \text{GeV}^{-1} \, \text{s}^{-1} \right)$, evaluated at the energies given by the array, $\texttt{energies}$. The result obtained using this function is a list of values of differential flux, $dN_{\gamma}/dE_{\gamma }$ $\left(\text{GeV}^{-1}\,\text{cm}^{-2}\, \text{s}^{-1}\right)$, evaluated at the energies given by the array, $\texttt{energies}$.

\item $\texttt{GCascadeDiffuseConstant[injected spectrum, max distance (z),}$ \\
$\texttt{luminosity function}$ $\rho \texttt{(z)]}$ is a function which produces the observed diffuse differential spectrum taking into account cosmological expansion and electromagnetic cascade evolution. $\texttt{GCascadeDiffuseConstant}$ uses the luminosity function as an input which must be formatted as list of ordered pairs, ($z$,$\rho$($z$)), of values of $z$ and comoving number density, $\rho$($z$). The values of $z$ can be any finite list of $z$ from 0 to 10. $\gamma$-Cascade provides two convenient lists of $z$, $\texttt{zReg}$ and $\texttt{diffuseDistances}$. $\texttt{zReg}$ is a list that contains values of $z$ from 0 to 10 in increments of 0.01. $\texttt{diffuseDistances}$ is a list that contains values of $z$ from 0 to 10 in progressively increasing step sizes. $\texttt{diffuseDistances}$ is the list of $z$-values used by the code to carry out the numerical integration (i.e. boundary of integration bins). This is a good list to use when $\rho$(z) is mainly localized near $z$=0. The injected differential spectrum should be formatted as an array of values of differential flux, $dN_{\gamma}/dE_{\gamma }$ $\left( \text{GeV}^{-1} \, \text{s}^{-1} \right)$, evaluated at the energies given by the array, $\texttt{energies}$. The result obtained using this function is a list of values of diffuse differential flux, $dN_{\gamma}/dE_{\gamma }$ $\left(\text{GeV}^{-1} \, \text{cm}^{-2} \, \text{s}^{-1}\, \text{sr}^{-1}\right)$, evaluated at the energies given by the array, $\texttt{energies}$.

\item $\texttt{GCascadeDiffuse[injected spectrum, max distance (z),}$ \\
$\texttt{luminosity function}$ $\rho \texttt{(z)]}$ is a function which produces the observed diffuse differential spectrum taking into account cosmological expansion and electromagnetic cascade evolution. $\texttt{GCascadeDiffuse}$ uses the luminosity function as an input which must be formatted as list of ordered pairs, ($z$,$\rho$($z$)), of values of $z$ and comoving number density, $\rho$($z$). The values of $z$ can be any finite list of $z$ from 0 to 10. $\gamma$-Cascade provides two convenient lists of $z$, $\texttt{zReg}$ and $\texttt{diffuseDistances}$. $\texttt{zReg}$ is a list that contains values of $z$ from 0 to 10 in increments of 0.01. $\texttt{diffuseDistances}$ is a list that contains values of $z$ from 0 to 10 in progressively increasing step sizes. $\texttt{diffuseDistances}$ is the list of $z$-values used by the code to carry out the numerical integration (i.e. boundary of integration bins). This is a good list to use when $\rho$(z) is mainly localized near $z$=0. Additionally, $\texttt{GCascadeDiffuse}$ uses an evolving differential spectrum, as an input. The injected spectrum must be an array of spectra, each formatted as an array of values of differential flux, $dN_{\gamma}/dE_{\gamma }$ $\left(\text{GeV}^{-1} \, \text{s}^{-1}\right)$, evaluated at the energies given by the array, $\texttt{energies}$. Each member of the injected spectrum array must be the spectrum injected at the z-values given by, $\texttt{diffuseDistances}$. The result obtained using this function is a list of values of diffuse differential flux, $dN_{\gamma}/dE_{\gamma }$ $\left(\text{GeV}^{-1} \, \text{cm}^{-2} \, \text{s}^{-1}\, \text{sr}^{-1}\right)$, evaluated at the energies given by the array, $\texttt{energies}$.
\end{itemize}

Computational time is approximately linear in max distance, z, for all functions listed above. A test computer running on 8Gb of RAM and a 2.3GHz Intel i5-6200U CPU was used to characterize the computational time using a flat power spectrum across the entire energy range. When $\texttt{GCascadeAttenuate}$, $\texttt{GCascadePoint}$, $\texttt{GCascadeDiffuseConstant}$, and $\texttt{GCascadeDiffuse}$ were run from a max distance of z = 0.1, the computational times were 2s, 85s, 87s, and 94s accordingly. It should be noted that the shape of the injected spectrum does not impact computational time significantly. 

Additionally, $\gamma$-Cascade allows the user to modify the field strength of the intergalactic magnetic field at $z=0$ using the following function:
\begin{itemize}
\item $\texttt{changeMagneticField[magnetic field, name]}$ is a function which changes the intergalactic magnetic field and produces new libraries. The first argument is the new magnetic field at $z=0$ in Gauss and the second is a string which will be prepended to the default $\texttt{libraryNames}$ string during export. The final name of each library will start with the $\texttt{name}$ argument making it simple to change $\texttt{libraryNames}$ in the package file by simply adding this string to the beginning. The user must import the package once again after running this function and modifying $\texttt{libraryNames}$ in the package file in order to have working functionality with the new magnetic field. It should be noted that changing the magnetic field to anything under about 0.1 nG will leave the calculation done by $\gamma$-Cascade largely unchanged and will only affect the highest energies, $E_{e}>10^{19}$ eV. Thus, in general, it is not recommended to change the nominal intergalactic magnetic field.
\end{itemize}

\begin{figure}
\centering
\includegraphics[width=0.8\textwidth]{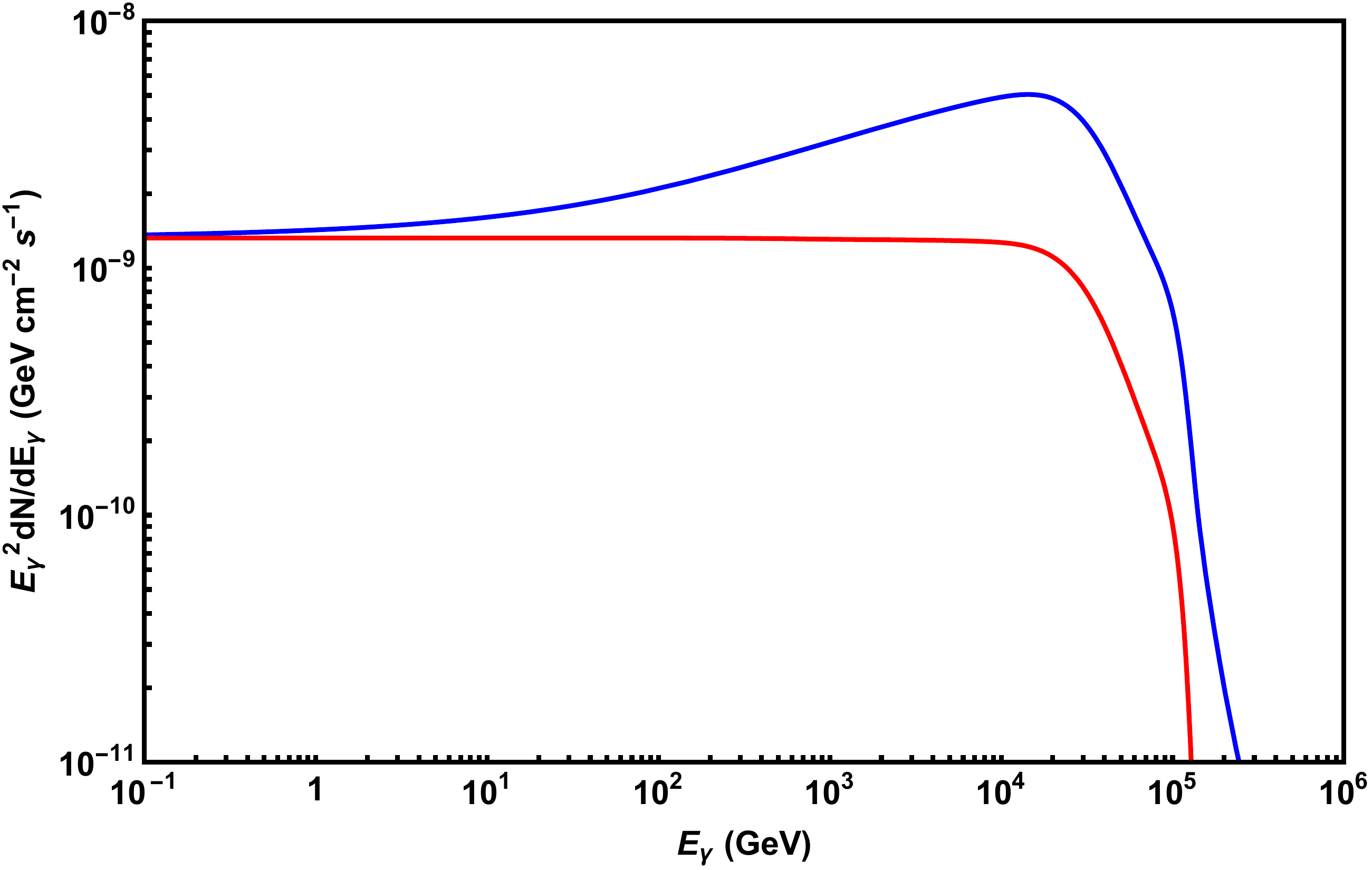}
\caption{The calculated spectra coming from a source with a flat spectrum, $dN_{\gamma}/dE_{\gamma } \propto E_{\gamma}^{-2}\exp{\left( -E_{\gamma}/(10^{13}\; \text{GeV})\right)}$ are shown above. This source was placed at $z=0.001$. The red curve is the attenuated spectrum obtained from $\texttt{GCascadeAttenuate}$. The blue curve is the spectrum after cascade development obtained using $\texttt{GCascadePoint}$.}
\label{Point}
\end{figure}

\begin{figure}
\centering
\includegraphics[width=0.8\textwidth]{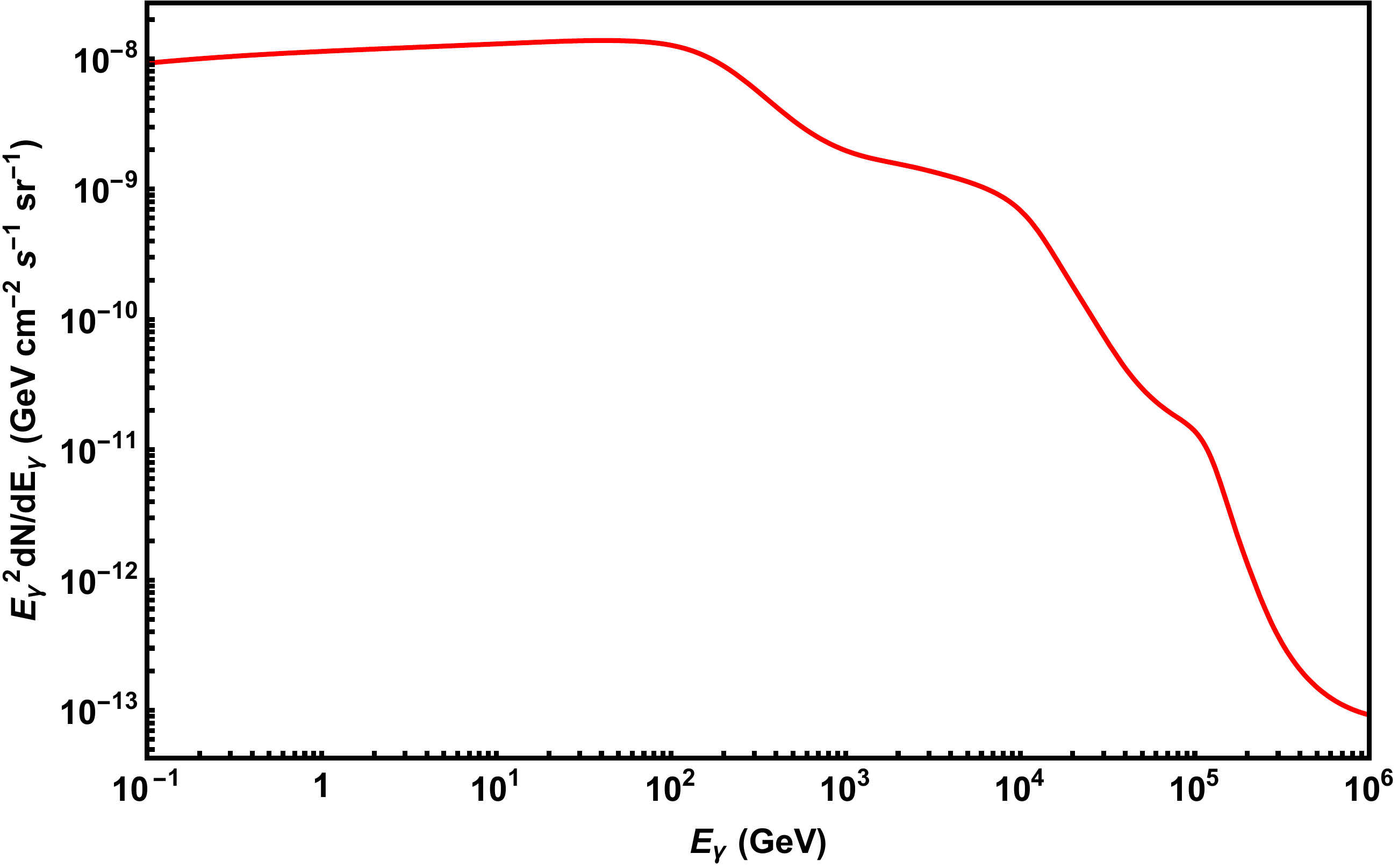}
\caption{The resulting diffuse spectrum obtained from running $\texttt{GCascadeDiffuse}$ on a class of sources with a flat spectrum, $dN_{\gamma}/dE_{\gamma } \propto E_{\gamma}^{-2}\exp{\left( -E_{\gamma}/(10^{13}\; \text{GeV})\right)}$ and a constant source density per comoving volume of $5\times 10^{-6}$ $\text{Mpc}^{-3}$ between $z=10^{-6}$ and $z=5$ is shown above.}
\label{Diffuse}
\end{figure}

%
As an illustrative example, Fig. \ref{Point} shows the resulting spectrum from the propagation of a hypothetical source with a flat, cutoff, spectrum given by the following:
\begin{equation}
dN_{\gamma}/dE_{\gamma } = 3.12\times10^{42}\left(\text{GeV}\;\text{s}^{-1}\right) E_\gamma^{-2} \exp{\left( -E_{\gamma}/(10^{13}\; \text{GeV})\right)},
\end{equation}
where the normalization is calculated for a source with an apparent isotropic luminosity above 250 GeV of $5\times10^{39}$ $\text{ergs}\;\text{s}^{-1}$. Additionally, the diffuse background generated by a distribution of such sources, with a constant source density per comoving volume of $5\times 10^{-6}$ $\text{Mpc}^{-3}$ between $z=10^{-6}$ and $z=5$, is shown in Fig. \ref{Diffuse}.

\section{Summary and Conclusions}
$\gamma$-Cascade is a Mathematica package for modeling gamma-ray propagation through cosmological distances taking into account electromagnetic cascade development, synchrotron cooling, and cosmological expansion. $\gamma$-Cascade allows users to calculate the observed gamma-ray flux from point sources as well as the observed diffuse flux from a distribution of sources. The code uses a semi-analytic model for electromagnetic interactions in order to calculate spectra in a fast, simple, and user-friendly manner. In the future, additional Mathematica code will be made available to modify the EBL model which will export the necessary modified $\gamma$-Cascade libraries. Users are encouraged to contact the author with any questions regarding $\gamma$-Cascade.
\bigskip

\textbf{Acknowledgments.} Special thanks to Dan Hooper and Sam McDermott for helpful discussions and for their help with the testing and debugging of the code as well as Adam Kline for his helpful discussions during the implementation of the code. CB is supported by the US National Science Foundation Graduate Research Fellowship under grant numbers DGE-1144082 and DGE-1746045.

\bibliography{gammaCascadeBib}
 \bibliographystyle{JHEP}

\end{document}